\documentclass[%
reprint, 
tightenlines, 
superscriptaddress, 
showpacs, 
nobibnotes, 
amsmath, amssymb, 
aps, 
pra,
floatfix, 
dvipdfmx
]{revtex4-2}

\usepackage{latexsym}  
\usepackage{graphicx}
\usepackage{dcolumn}
\usepackage{bm}
\usepackage[usenames]{color} 

\allowdisplaybreaks

\begin{document}

\preprint{APS/123-QED}

\title{su(N) Mermin-Ho relation}

\author{Emi Yukawa} 
\email{A29529@rs.tus.ac.jp} 
\affiliation{RIKEN Center for Emergent Matter Science, 2-1, Hirosawa, Wako-shi, Saitama, 351-0198, Japan} 
\affiliation{Department of Physics, Faculty of Science Devision I, Tokyo University of Science, 1-3, Kagurazaka, Shinjuku-ku, Tokyo, 
162-8602, Japan }

\date{\today}

\begin{abstract}
The Mermin-Ho relation expresses the vorticity of a coreless su($2$) vortex in terms of the spin-$1$ or $l$ vector 
which characterizes fully polarized superfluid textures. 
We generalize it to an su($N$) vortex which is applicable to arbitrarily polarized superfluid textures with higher spin or angular momentum. 
The obtained relation is expressed in terms of the mean-field generators and their structure factors. 
\end{abstract}

\maketitle

\section{\label{sec:1}Introduction} 
The Mermin-Ho relation was originally discussed in the context of a superfluid $^3$He~\cite{MerminHo} and expresses
the vorticity of a nonsingular vortex in terms of the texture of the superfluid order parameter as  
\begin{align} 
	\nabla \times \bm{v} = \frac{\hbar}{2M} \sum_{\mu , \nu , \lambda = x, y, z} {\epsilon}_{\mu \nu \lambda} 
	l_{\mu} (\nabla l_{\nu} ) \times (\nabla l_{\lambda} ), \label{eq:MerminHo}
\end{align} 
where $\bm{v}$ is the velocity field, $l_{\mu}$ ($\mu = x, y, z$) is the $\mu$ component of the $\bm{l}$ vector, and ${\epsilon}_{\mu \nu \lambda}$ is the rank-$3$ antisymmetric tensor. 
Reflecting the so($3$) symmetry of the order parameter, Eq.~(\ref{eq:MerminHo}) involves the so($3$) generators $l_{\mu}$ and 
their structure factors ${\epsilon}_{\mu \nu \lambda}$. 
In a ferromagnetic spin-$F$ Bose-Einstein condensate (BEC), the multicomponent order parameter also has the so($3$) symmetry. 
Here the Mermin-Ho relation relates the velocity field of a coreless vortex to the unit vector $\bm{n}$ parallel to the local 
magnetization vector as~\cite{Lamacraft1, Lamacraft2, KudoKawaguchi}
\begin{align} 
	\nabla \times \bm{v} = \frac{\hbar F}{2M} \sum_{\mu ,\nu ,\lambda = x, y, z} {\epsilon}_{\mu \nu \lambda} 
	n_{\mu} (\nabla n_{\nu} ) \times (\nabla n_{\lambda}). \label{eq:spinorMerminHo_ferro}
\end{align} 
Here, the velocity field $\bm{v}$ is defined in terms of the 
multicomponent order parameter ${\psi}_m$ ($m = 1, 2 \cdots , N; \ N \equiv 2F+1$) as 
\begin{align} 
	\bm{v} = \frac{\hbar}{2M\rho i} \sum_{m=1}^N [{\psi}_m^* (\nabla {\psi}_m ) - (\nabla {\psi}_m^* ) {\psi}_m ], \label{eq:velocity}
\end{align} 
where $\rho \equiv \sum_{m=1}^N | {\psi}_m |^2$ is the local density of particle number. 

More generally, the multicomponent order parameter has the su($N$) symmetry and it cannot be expressed in terms of 
the magnetization vector alone. 
In fact, for a spin-$1$ BEC, the vorticity is given by
\begin{align} 
	\nabla \times \bm{v} = \frac{\hbar}{8M} \sum_{j,k,l = 1}^8 C^{(3)}_{jkl} {\lambda}^{(3)}_j (\nabla {\lambda}^{(3)}_k ) 
	\times (\nabla {\lambda}^{(3)}_l ), \label{eq:spin1MerminHo_GellMann}
\end{align} 
where $C^{(N)}_{jkl}$ is the su($N$) structure factor and ${\lambda}^{(N)}_j$ is the expectation value of the irreducible 
representation of the $j$th generator ${\Lambda}^{(N)}_j$ defined by 
\begin{align} 
	{\lambda}^{(N)}_j \equiv \sum_{m,n =1}^{N} ({\Lambda}^{(N)}_j)_{mn} {\xi}_m^* {\xi}_n, \label{eq:spin1lambda}
\end{align} 
with the multicomponent order parameter ${\xi}_m$ normalized to unity as ${\xi}_m \equiv {\psi}_m / \sqrt{\rho}$~\cite{Yukawa}. 

In this paper, we generalize this result to an arbitrary spin-$F$ BEC and derive an explicit expression of the vorticity in terms of the su($N$) 
generators and their structure factors. This is a natural extension of Eqs.~(\ref{eq:MerminHo}) and (\ref{eq:spin1MerminHo_GellMann}) and 
applicable to arbitrarily polarized spin-$F$ textures. 

\section{\label{sec:2}su(N) generators and structure factors} 
In this section, we introduce matrix representations of su($N$) generators and their structure factors, which we will use 
in deriving the su($N$) Mermin-Ho relation in the next section. 
An irreducible representation of any su($N$) generator can be expressed in terms of an $N$-dimensional matrix. 
Let the matrices of the su($N$) generators be ${\Lambda}_j^{(N)}$ ($j = 1, 2, \cdots , N^2 - 1$), which are normalized so that 
their Frobenius norms are equal to $\sqrt{2}$. 
We define 
\begin{align} 
	& {\Lambda}_{(n - 1)^2 + 2 (m - 1)}^{(N)} \equiv X_{mn}^{(N)}, \label{eq:lambda_X} \\
	& {\Lambda}_{(n - 1)^2 + 2m - 1}^{(N)} \equiv Y_{mn}^{(N)}, \label{eq:lambda_Y} \\
	& {\Lambda}_{n^2 - 1}^{(N)} \equiv \sqrt{\frac{2}{n (n-1)}} \mathrm{diag} 
	(\underbrace{1, \cdots , 1}_{n-1}, -n+1) \oplus O^{(N-n)}, \label{eq:lambda_diag}
\end{align}  
where $m$ and $n$ satisfy $1 \leq m < n \leq N$, the $m^{\prime}n^{\prime}$-components of the matrices $X_{mn}^{(N)}$ and 
$Y_{mn}^{(N)}$ are given by 
\begin{align} 
	&(X_{mn}^{(N)})_{m^{\prime}n^{\prime}} \equiv {\delta}_{mm^{\prime}} {\delta}_{nn^{\prime}} 
	+  {\delta}_{mn^{\prime}} {\delta}_{nm^{\prime}},  \label{eq:Pauli_X} \\
	&(Y_{mn}^{(N)})_{m^{\prime}n^{\prime}} \equiv -i {\delta}_{mm^{\prime}} {\delta}_{nn^{\prime}} 
	+  i {\delta}_{mn^{\prime}} {\delta}_{nm^{\prime}}, \label{eq:Pauli_Y}
\end{align} 
and $O^{(N-n)}$ denotes the $(N-n)$-dimensional zero matrix. 
Equations~(\ref{eq:lambda_X})-(\ref{eq:lambda_diag}) imply that the first $N(N-2)$ matrices of su($N$) generators can be expressed 
recursively in terms of su($N-1$) generators as ${\Lambda}_j^{(N-1)} \oplus O^{(1)}$ ($j = 1, 2, \cdots , N(N-2)$). 
Here, we note that the matrices given in Eq.~(\ref{eq:lambda_diag}), which are diagonal, altogether form the Cartan subalgebra and their 
linear combination give $Z_{mn}^{(N)}$ ($1 \leq m < n \leq N$) whose $m^{\prime} n^{\prime}$ component is expressed as  
\begin{align} 
	(Z_{mn}^{(N)})_{m^{\prime} n^{\prime}} \equiv {\delta}_{mm^{\prime}} - {\delta}_{nn^{\prime}}. \label{eq:Pauli_Z}
\end{align} 

The structure factor $C_{jkl}^{(N)}$ is an antisymmetric tensor with respect to $j$, $k$, and $l$, and satisfies the following 
commutation relation: 
\begin{align} 
	[ {\Lambda}_j^{(N)}, {\Lambda}_k^{(N)} ] = i \sum_{l=1}^{N^2-1} C_{jkl}^{(N)} {\Lambda}_l^{(N)}. \label{eq:comm_relation}
\end{align} 
It follows from Eqs.~(\ref{eq:lambda_X})-(\ref{eq:lambda_diag}) that any nonzero structure factor is classified into the 
following three types according to the commutation relation involving the structure factor. 

The first type is constituted from structure factors of su($N-1$) commutation relations, that is, 
\begin{align} 
	[ {\Lambda}_j^{(N)}, {\Lambda}_k^{(N)} ] = i \sum_{l=1}^{N(N-2)} C_{jkl}^{(N-1)} {\Lambda}_l^{(N)}, \label{eq:comm_suN-1}
\end{align} 
with $1 \leq j, k, l \leq N(N-2)$, which implies that $C_{jkl}^{(N)} = C_{jkl}^{(N-1)}$. 

The second type is comprised of structure factors of the following set of commutation relations: 
\begin{align} 
	[ {\Lambda}_j^{(N)}, {\Lambda}_k^{(N)} ] = i \sum_{n=2}^N C_{jkn^2-1}^{(N)} {\Lambda}_{n^2-1}^{(N)}, \label{eq:comm_S2}
\end{align} 
where $j$ and $k$ are given by 
\begin{align} 
	&j = (N-1)^2 + 2(m - 1), \label{eq:S2-1} \\ 
	&k = (N - 1)^2 + 2m - 1, \label{eq:S2-2}
\end{align} 
with $1 \leq m < N$. 
From Eqs.~(\ref{eq:comm_S2})-(\ref{eq:S2-2}), ${\Lambda}_j^{(N)}$ and ${\Lambda}_k^{(N)}$ in Eq.~(\ref{eq:comm_S2}) are equivalent to 
\begin{align} 
	{\Lambda}_j^{(N)} = X_{mN}^{(N)}, \ {\Lambda}_k^{(N)} = Y_{mN}^{(N)}, 
\end{align} 
and 
\begin{align} 
	\sum_{n=2}^N C_{jkn^2-1}^{(N)} {\Lambda}_{n^2-1}^{(N)} = 2 Z_{mN}^{(N)}, \label{eq:Zmn}
\end{align} 
which implies that the commutation relation in Eq.~(\ref{eq:comm_S2}) can be simplified to 
\begin{align} 
	[ X_{mN}^{(N)}, Y_{mN}^{(N)} ] = 2i Z_{mN}^{(N)}. \label{eq:comm_XYZ}
\end{align} 
The structure coefficients $C_{jkl}^{(N)}$'s are given in the Appendix~\ref{asec:1}. 
Here, we note that $C_{jkl}^{(N)}$ is antisymmetric with respect to $j$, $k$ and $l$. 

The third type involves structure factors $|C_{jkl}^{(N)}| = 1$ and commutation relations with them are given by 
\begin{align}
	[ {\Lambda}_j^{(N)}, {\Lambda}_k^{(N)} ] = i {\Lambda}_l^{(N)}, \label{eq:comm_S3} 
\end{align} 
where $(j, k,l)$ is given by either of 
\begin{widetext}
\begin{align} 
	(j, k, l) = & ( (n-1)^2 + 2(m-1), (N-1)^2 + 2(m-1), (N-1)^2 + 2n-1 ), \label{eq:S3-1} \\ 
	& ( (n-1)^2 + 2(m-1), (N-1)^2 + 2(n-1), (N-1)^2 + 2m-1 ), \label{eq:S3-2} \\ 
	&( (n-1)^2 + 2m-1, (N-1)^2 + 2(m-1), (N-1)^2 + 2(n-1) ),  \label{eq:S3-3} \\ 
	&\text{or } ( (n-1)^2 + 2m-1, (N-1)^2 + 2m-1, (N-1)^2 + 2n-1), \label{eq:S3-4}
\end{align} 
\end{widetext} 
with $1 \leq m < n < N$. 
This implies that the commutation relations corresponding to Eqs.~(\ref{eq:S3-1})-(\ref{eq:S3-4}) can be expressed in terms of 
$X_{mn}^{(N)}$ and $Y_{mn}^{(N)}$ in Eqs.~(\ref{eq:Pauli_X}) and (\ref{eq:Pauli_Y}) as 
\begin{align} 
	& [X_{mn}^{(N)}, X_{mN}^{(N)} ] = i Y_{nN}^{(N)}, \label{eq:comm_mnN1} \\
	& [X_{mn}^{(N)}, X_{nN}^{(N)}] = i Y_{mN}^{(N)}, \label{eq:comm_mnN2} \\
	& [Y_{mn}^{(N)}, X_{mN}^{(N)} ] = i X_{nN}^{(N)}, \label{eq:comm_mnN3}\\ 
	& [Y_{mn}^{(N)}, Y_{mN}^{(N)} ] = i Y_{nN}^{(N)}. \label{eq:comm_mnN4}
\end{align} 

\section{\label{sec:3}Generalized Mermin-Ho relation} 
The definition of the velocity field given in Eq.~(\ref{eq:velocity}) implies that the vorticity can be expressed as 
\begin{align} 
	\nabla \times \bm{v} = \frac{i\hbar}{M} \sum_{m,n,n^{\prime} = 1}^N a_{mn} (\nabla a_{nn^{\prime}}) 
	\times (\nabla a_{n^{\prime}m} ), \label{eq:suNMerminHo'}
\end{align} 
where $a_{mn} \equiv {\xi}_m^* {\xi}_n$. 
Equation~(\ref{eq:suNMerminHo'}) is shown in Appendix~\ref{asec:2}. 
By mathematical induction, we can prove that Eq.~(\ref{eq:suNMerminHo'}) is equivalent to 
\begin{align} 
	\nabla \times \bm{v} = \frac{\hbar}{8M} \sum_{j,k,l=1}^{N^2-1} C_{jkl}^{(N)} {\lambda}_j^{(N)} (\nabla {\lambda}_k^{(N)} ) \times (\nabla 
	{\lambda}_l^{(N)} ), \label{eq:suNMerminHo}
\end{align} 
which is the main result of this paper. 

The outline of the proof is as follows. 
First, the sum on the right-hand side of Eq.~(\ref{eq:suNMerminHo}) is rewritten as a sum over a set of subscripts $(j,k,l)$ 
with a nonzero structure factor $C_{jkl}^{(N)} \neq 0$ as 
\begin{align} 
	\sum_{j,k,l=1}^{N^2-1} \cdots = \sum_{(j,k,l) \in S^{(N)}} \cdots ,  \label{eq:sum}
\end{align} 
where $S^{(N)}$ is defined by 
\begin{align} 
	S^{(N)} = \{ (j,k,l); \ ^{\forall}j,k,l = 1, 2, \cdots , N^2 - 1, \ C_{jkl}^{(N)} \neq 0 \}. \label{eq:S}
\end{align} 
As shown in Sec.~\ref{sec:2}, a nonzero structure factor is classified into the three types; hence $S^{(N)}$ can be decomposed into 
three subsets $S_1^{(N)}$, $S_2^{(N)}$, and $S_3^{(N)}$ corresponding to the three types of $C_{jkl}^{(N)}$. 
It follows from Eqs.~(\ref{eq:comm_suN-1}), (\ref{eq:comm_XYZ}), and (\ref{eq:comm_S3}) that $S_1^{(N)}$, $S_2^{(N)}$, and $S_3^{(N)}$ can be expressed respectively as  
\begin{align} 
	S_1^{(N)} \equiv \{ (j,k,l); \ & ^{\forall}j,k,l = 1, 2, \cdots , N(N-2), \nonumber \\ 
	& C_{jkl}^{(N)} = C_{jkl}^{(N-1)} \} , \label{eq:S1}
\end{align} 
\begin{align} 
	S_2^{(N)} \equiv \{ (j,k,l); \ & \text{$j$, $k$, $l = n^2$ ($1 < n \leq N$)} \nonumber \\ 
	&\text{given in Eq.~(\ref{eq:S2-1})-(\ref{eq:S2-2})} \nonumber \\ 
	& \text{and their permutations} \} , \label{eq:S2}
\end{align} 
and 
\begin{align} 
	S_3^{(N)} \equiv \{ (j,k,l); \ & \text{$j$, $k$, $l = n^2 -1$ ($n = 2, \cdots , N$)} \nonumber \\ 
	& \text{given in Eq.~(\ref{eq:S3-1})-(\ref{eq:S3-4})} \nonumber \\ 
	& \text{and their permutations} \} . \label{eq:S3}
\end{align} 

Then, in the inductive hypothesis, we assume that Eq.~(\ref{eq:suNMerminHo}) holds for the su($N-1$) case. 
\begin{align} 
	&\sum_{(j,k,l) \in S_1^{(N)}} C_{jkl}^{(N)} {\lambda}_j^{(N)} (\nabla {\lambda}_k^{(N)} ) \times (\nabla {\lambda}_l^{(N)} ) 
	\nonumber \\ 
	&= 8i \sum_{m ,n ,n^{\prime} = 1}^{N-1} a_{mn} (\nabla a_{nn^{\prime} } ) \times (\nabla a_{n^{\prime} m} ). \label{eq:s1}
\end{align} 
The sum over $S_2^{(N)}$ can be obtained as 
\begin{align} 
	&\sum_{(j,k,l) \in S_2^{(N)}} C_{jkl}^{(N)} {\lambda}_j^{(N)} (\nabla {\lambda}_k^{(N)} ) \times (\nabla {\lambda}_l^{(N)} ) \nonumber \\
	&= 8i \sum_{m=1}^N \{ a_{mN} [ (\nabla a_{Nm} ) \times (\nabla a_{mm} ) \nonumber \\
	&+ (\nabla a_{NN} ) \times (\nabla a_{Nm} ) ] \nonumber \\
	&+ a_{Nm} [ (\nabla a_{mm} ) \times (\nabla a_{mN} ) + (\nabla a_{mN} ) \times (\nabla a_{NN} ) ] \nonumber \\ 
	&+ a_{mm} (\nabla a_{mN} ) \times (\nabla a_{Nm} ) + a_{NN} (\nabla a_{Nm} ) \times (\nabla a_{mN} ) \}. 
	\label{eq:s2}
\end{align} 
The sum over $S_3^{(N)}$ is given by 
\begin{align} 
	&\sum_{(j,k,l) \in S_3^{(N)}} C_{jkl}^{(N)} {\lambda}_j^{(N)} (\nabla {\lambda}_k^{(N)} ) \times (\nabla {\lambda}_l^{(N)} ) \nonumber \\ 
	&= 8i \sum_{1 \leq m < n < N} [ a_{mn} (\nabla a_{nN} ) \times (\nabla a_{Nm}) \nonumber \\
	&+ a_{nm} (\nabla a_{mN} ) \times (\nabla a_{Nn} ) + a_{Nm} (\nabla a_{mn} ) \times (\nabla a_{nN} ) \nonumber \\
	&+ a_{Nn} (\nabla a_{nm} ) \times (\nabla a_{mN} ) + a_{nN} (\nabla a_{Nm} ) \times (\nabla a_{mn} ) \nonumber \\ 
	&+ a_{mN} (\nabla a_{Nn} ) \times (\nabla a_{nm} ) ]. \label{eq:s3}
\end{align} 
Taking sum of Eqs.~(\ref{eq:s1})-(\ref{eq:s3}), we obtain 
\begin{align} 
	&\sum_{(j,k,l) \in S^{(N)}} C_{jkl}^{(N)} {\lambda}_j^{(N)} (\nabla {\lambda}_k^{(N)} ) \times (\nabla {\lambda}_l^{(N)} ) 
	\nonumber \\ 
	&= 8i \sum_{m,n,n^{\prime} = 1}^N a_{mn} (\nabla a_{nn^{\prime}} ) \times (\nabla a_{n^{\prime} m} ), 
\end{align} 
which implies that Eq.~(\ref{eq:suNMerminHo}) holds for $N$. 
Then by the principle of the mathematical induction, Eq.~(\ref{eq:suNMerminHo}) is proved for any spin-degrees of freedom. 
The proof of Eqs.~(\ref{eq:s2}) and (\ref{eq:s3}) is shown in Appendix~\ref{asec:3}.  

\section{\label{sec:5} Conclusion} 
The original Mermin-Ho relation is applicable to the fully-polarized spin case, where the underlying algebra is su($2$). 
We have generalized it to arbitrarily polarized cases of the general spin degrees of freedom with the su($N$) algebra. 
The main result is given by Eq.~(\ref{eq:suNMerminHo}), where the su($N$) generators are given in 
Eqs.~(\ref{eq:lambda_X})-(\ref{eq:lambda_diag}) and the nonzero structure factors are categorized into three types 
corresponding to the three different types of commutation relations in Eqs.~(\ref{eq:comm_suN-1}), (\ref{eq:comm_S2}), 
and (\ref{eq:comm_S3}). 
We expect the obtained result can be applied to high spin and high angular momentum systems. 

\begin{acknowledgements} 
The author thank Masahito Ueda for fruitful discussions. 
\end{acknowledgements} 

\appendix 
\section{\label{asec:1} Derivation of the coefficients in Eq.~(\ref{eq:comm_XYZ})} 
We solve Eq.~(\ref{eq:Zmn}) to obtain the structure factors in Eq.~(\ref{eq:comm_XYZ}) in two cases, i.e., $m = 1$ and 
$1 < m < N$. 
For the sake of simplicity, let we define 
\begin{align} 
	{\alpha}_{mn} \equiv \frac{C_{jkn^2-1}}{\sqrt{2n(n-1)}} 
\end{align} 
with $j$ and $k$ given in Eqs~(\ref{eq:S2-1}) and (\ref{eq:S2-2}). 

In the case of $m = 1$, it follows from Eqs.~(\ref{eq:lambda_diag}), (\ref{eq:Pauli_Z}), and the second equality of 
Eq.~(\ref{eq:comm_XYZ}) that 
\begin{align} 
	{\alpha}_{12} + {\alpha}_{13} + \cdots + {\alpha}_{1N} &= 1, \label{eq:proofm=1-1} \\ 
	- (n-1) {\alpha}_{1n} + \sum_{n^{\prime}=n+1}^N {\alpha}_{1n^{\prime}} &= 0 \ 
	(1 < n < N), \label{eq:proofm=1-2} \\
	- (N - 1) {\alpha}_{1N} &= -1. \label{eq:proofm=1-3}
\end{align} 
Equation~(\ref{eq:proofm=1-3}) gives 
\begin{align} 
	{\alpha}_{1N} = \frac{1}{N-1}, \label{eq:proofm=1-4}
\end{align} 
Substituting Eq.~(\ref{eq:proofm=1-4}) in ${\alpha}_{1N}$ of Eq.~(\ref{eq:proofm=1-2}) with $n=N-1$, we obtain 
\begin{align} 
	{\alpha}_{1N-1} = \frac{1}{(N-1)(N-2)}. \label{eq:proofm=1-5}
\end{align} 
By repeating this procedure in descending order of $n$ down to $n = 2$, we have 
\begin{align} 
	{\alpha}_{1n} = \frac{1}{n(n-1)}. \label{eq:proofm=1-6}
\end{align} 
We can confirm that Eq.~(\ref{eq:proofm=1-1}) holds for the solutions given in Eqs.~(\ref{eq:proofm=1-4})-(\ref{eq:proofm=1-6}). 
Therefore, the solution of Eq.~(\ref{eq:Zmn}) for $m=1$ is given by 
\begin{align} 
	C_{(N-1)^2,(N-1)^2+1,n^2-1}^{(N)} = \left \{ \begin{array}{ll} 
	\sqrt{\frac{2}{n(n-1)}} & (1 < n < N); \\
	\sqrt{\frac{2N}{N-1}} & (n=N). 
	\end{array} \right. \label{eq:proofm=1-7} 
\end{align} 

In the case of $1 < m < N$, Eqs.~(\ref{eq:lambda_diag}), (\ref{eq:Pauli_Z}), and the second equality in Eq.~(\ref{eq:Zmn}) are 
equivalent to the following linear equations:  
\begin{align} 
	{\alpha}_{m2} + {\alpha}_{m3} + \cdots + {\alpha}_{mN} &= 0, \label{eq:proofm>1-1} \\ 
	- (n-1) {\alpha}_{mn} + \sum_{n^{\prime}=n+1}^N {\alpha}_{mn^{\prime}} &= {\delta}_{mn} \
	(1 < n < N), \label{eq:proofm>1-2} \\
	- (N - 1) {\alpha}_{mN} &= -1. \label{eq:proofm>1-3}
\end{align} 
By applying the same procedure as in the case of $m=1$ to Eq.~(\ref{eq:proofm>1-1})-(\ref{eq:proofm>1-3}), 
we obtain 
\begin{align} 
	C_{jkn^2-1}^{(N)} = \left \{ \begin{array}{ll} 
	0 & (n < m); \\ 
	- \sqrt{\frac{2(m-1)}{m}} & (n = m); \\ 
	\sqrt{\frac{2}{n(n-1)}} & (m < n < N); \\
	\sqrt{\frac{2N}{N-1}} & (n = N), 
	\end{array} \right. \label{eq:proofm>1-4}
\end{align} 
with $j$ and $k$ given by Eqs.~(\ref{eq:S2-1}) and (\ref{eq:S2-2}). 
Equations~(\ref{eq:proofm=1-7}) and (\ref{eq:proofm>1-4}) imply that the coefficients ${\alpha}_{mn}$ in Eq.~(\ref{eq:comm_XYZ}) 
can be expressed as in Eq.~(\ref{eq:proofm>1-4}) regardless of the value of $m$, provided that $n > 1$. 

\section{\label{asec:2} Proof of Eq.~(\ref{eq:suNMerminHo'})} 
Substituting ${\psi}_m = \sqrt{\rho} {\xi}_m$ in Eq.~(\ref{eq:velocity}), we obtain 
\begin{align} 
	\bm{v} &= \frac{\hbar}{2M\rho i} \sum_{m=1}^N \left [ \sqrt{\rho} {\xi}_m^* (\nabla  \sqrt{\rho} {\xi}_m )
	- \mathrm{h.c.} \right ] \nonumber \\ 
	&= \frac{\hbar}{2M i} \sum_{m=1}^N \left [ {\xi}_m^* (\nabla {\xi}_m ) - (\nabla {\xi}_m^* ) {\xi}_m \right ] 
	\label{eq:velocity'} \\ 
	&= \frac{\hbar}{M i} \sum_{m=1}^N {\xi}_m^* (\nabla {\xi}_m ) 
	= - \frac{\hbar}{Mi} \sum_{m=1}^N (\nabla {\xi}_m^* ) {\xi}_m , \label{eq:velocity''}
\end{align} 
where the third and fourth equalities hold due to the identity $\nabla \sum_{m=1}^N | {\xi}_m|^2 = 0$. 
By taking the rotation of the first equality in Eq.~(\ref{eq:velocity'}), its $\mu$ ($\mu = x, y,$ or $z$) component becomes 
\begin{widetext} 
\begin{align} 
	(\nabla \times \bm{v})_{\mu} 
	&= \frac{\hbar}{Mi} \sum_{\nu , \lambda = x, y, z} \sum_{m=1}^N {\epsilon}_{\mu \nu \lambda} 
	( {\partial}_{\nu} {\xi}_m^*) ( {\partial}_{\lambda} {\xi}_m ) 
	= \frac{\hbar}{Mi} \sum_{\nu , \lambda = x, y, z} \sum_{m,n,n^{\prime}=1}^N {\epsilon}_{\mu \nu \lambda} 
	( {\partial}_{\nu} {\xi}_m^* {\xi}_n^* {\xi}_n ) ( {\partial}_{\lambda} {\xi}_m {\xi}_{n^{\prime}}^* {\xi}_{n^{\prime}}) \nonumber \\ 
	&= \frac{\hbar}{Mi} \sum_{\nu , \lambda = x, y, z} {\epsilon}_{\mu \nu \lambda} \left \{ \sum_{m,n,n^{\prime} =1}^N 
	( {\partial}_{\lambda} a_{n^{\prime}m} ) ( {\partial}_{\nu} a_{mn} ) a_{nn^{\prime}} 
	+ \sum_{m,n=1}^N {\xi}_m^* ( {\partial}_{\lambda} {\xi}_m) ( {\partial}_{\nu} {\xi}_n^* ) {\xi}_n \right \} \nonumber \\ 
	&=  \frac{\hbar}{Mi} \sum_{m,n,n^{\prime} =1}^N a_{mn} [(\nabla a_{nn^{\prime}} ) \times (\nabla a_{n^{\prime}m} )]_{\mu}
	- \frac{Mi}{\hbar} (\bm{v} \times \bm{v})_{\mu} = \text{right-hand side of Eq.~(\ref{eq:suNMerminHo'})}, \label{eq:rot_velocity}
\end{align} 
\end{widetext} 
where the identity $\sum_{m=1}^N |{\xi}_m|^2 = 1$ is used to obtain the right-hand side of the second equality and 
the expressions of $\bm{v}$ in Eq.~(\ref{eq:velocity''}) are substituted for $\sum_{m=1}^N {\xi}_m^* (\nabla {\xi}_m)$ and 
its complex conjugate on the right-hand side of the fourth equality.   

\section{\label{asec:3}Proof of Eqs.~(\ref{eq:s2}) and (\ref{eq:s3})} 
First, we prove Eq.~(\ref{eq:s2}). 
The antisymmetric properties of the structure factors and the outer products imply that the sum in Eq.~(\ref{eq:s2}) can be 
simplified as $\sum_{(j,k,l) \in S_2^{(N)} } = 2 \sum_{(j,k,l) \in S_2^{\prime (N)} }$, where $S_2^{\prime (N)}$ is the even-permutation 
subset of $S_2^{(N)}$ defined in Eq.~(\ref{eq:S2}). 
The expectation values ${\lambda}_{(N-1)^2 +2(m-1)}^{(N)}$ and ${\lambda}_{(N-1)^2 + 2m-1}^{(N)}$ are given by 
\begin{align} 
	&{\lambda}_{(N-1)^2 +2(m-1)}^{(N)} = a_{mN} + a_{Nm}, \\
	&{\lambda}_{(N-1)^2 +2m-1}^{(N)} = i (- a_{mN} + a_{Nm}),  
\end{align} 
since ${\Lambda}_{(n-1)^2 +2(m-1)}^{(N)} = X_{mn}^{(N)}$ and ${\Lambda}_{(n-1)^2 +2m-1}^{(N)} = Y_{mn}^{(N)}$. 
This implies that 
\begin{align} 
	&\sum_{n = m}^N C_{(N-1)^2+2(m-1),(N-1)^2+2m-1,n^2-1}^{(N)} {\lambda}_{n^2-1}^{(N)} \nonumber \\ 
	&= 2 \sum_{m^{\prime},n^{\prime} = 1}^N (Z_{mN}^{(N)})_{m^{\prime}n^{\prime}} a_{m^{\prime}n^{\prime}} 
	= 2 (a_{mm} - a_{NN}).
\end{align} 
Thus, the sum over $S_2^{(N)}$ can be expressed in terms of the $a_{mn}$ as 
\begin{widetext} 
\begin{align} 
	&\sum_{(j,k,l) \in S_2^{(N)} } C_{jkl}^{(N)} {\lambda}_j^{(N)} (\nabla {\lambda}_k^{(N)} ) \times (\nabla {\lambda}_l^{(N)} ) 
	= 2 \sum_{(j,k,l) \in S_2^{\prime (N)} } C_{jkl}^{(N)} {\lambda}_j^{(N)} (\nabla {\lambda}_k^{(N)} ) \times 
	(\nabla {\lambda}_l^{(N)} ) \nonumber \\ 
	=& 4i \sum_{m=1}^{N-1} \{ ( a_{mN} + a_{Nm} ) [ \nabla (- a_{mN} + a_{Nm}) ] \times [ \nabla (a_{mm} - a_{NN} ) ] 
	+ (- a_{mN} + a_{Nm} ) [ \nabla (a_{mm} - a_{NN} ) ] \times [ \nabla (a_{mN} + a_{Nm} ) ] \nonumber \\ 
	&+ (a_{mm} - a_{NN} ) [ \nabla ( a_{mN} + a_{Nm} ) ] \times [ \nabla ( - a_{mN} + a_{Nm} ) ] \} \nonumber \\ 
	=& 8i \sum_{m = 1}^N \{ a_{mN} [ (\nabla a_{Nm} ) \times (\nabla a_{mm} ) + (\nabla a_{NN} ) \times ( \nabla a_{Nm} ) ] 
	+ a_{Nm} [ (\nabla a_{mm} ) \times (\nabla a_{mN} ) + (\nabla a_{mN} ) \times ( \nabla a_{NN} ) ] \nonumber \\ 
	&+ a_{mm} (\nabla a_{mN} ) \times (\nabla a_{Nm} ) + a_{NN} (\nabla a_{Nm} ) \times (\nabla a_{mN} ) \}, 
\end{align} 
\end{widetext} 
where the sum on the right-hand side of the last equality is taken from $m = 1$ to $N$, since the terms are always zero for $m = N$. 

Next, we derive Eq.~(\ref{eq:s3}). 
In a manner similar to the proof of Eq.~(\ref{eq:s2}), we define the even-permutation subset of $S_3^{(N)}$ in Eq.~(\ref{eq:S3}) 
as $S_3^{\prime (N)}$. 
Since 
\begin{align} 
	&{\lambda}_{(n-1)^2+2(m-1)}^{(N)} = a_{mn} + a_{nm}, \\
	&{\lambda}_{(n-1)^2+2m-1}^{(N)} = i(-a_{mn} + a_{nm}),  
\end{align} 
we have 
\begin{widetext}
\begin{align} 
	&\sum_{(j,k,l) \in S_3^{(N)} } C_{jkl}^{(N)} {\lambda}_j^{(N)} (\nabla {\lambda}_k^{(N)} ) \times (\nabla {\lambda}_l^{(N)} ) 
	= 2 \sum_{(j,k,l) \in S_3^{\prime (N)} } C_{jkl}^{(N)} {\lambda}_j^{(N)} (\nabla {\lambda}_k^{(N)} ) \times (\nabla {\lambda}_l^{(N)} ) 
	\nonumber \\ 
	=& 2i \sum_{m=1}^{N-2} \sum_{n=m+1}^{N-1} \bigl \{ (a_{mn} + a_{nm}) [ \nabla (a_{mN} + a_{Nm} ) ] 
	\times [\nabla (- a_{nN} + a_{Nn} ) ] \nonumber \\ 
	&+ (a_{mn} + a_{nm}) [ \nabla (a_{nN} + a_{Nn} ) ] \times [\nabla (- a_{mN} + a_{Nm} ) ] \nonumber \\
	&+ (- a_{mn} + a_{nm}) [ \nabla (a_{mN} + a_{Nm} ) ] \times [\nabla (a_{nN} + a_{nN} ) ] \nonumber \\
	&+ (- a_{mn} + a_{nm}) [ \nabla (- a_{mN} + a_{Nm} ) ] \times [\nabla (- a_{nN} + a_{nN} ) ] \nonumber \\ 
	&+ (\text{cyclically permuted terms with respect to $m$, $n$, and $N$}) \bigl \} \nonumber \\
	=& 4i \sum_{m=1}^{N-2} \sum_{n=m+1}^{N-1} \bigl \{ (a_{mn} + a_{nm}) [ (\nabla a_{mN} ) \times (\nabla a_{Nn} ) 
	+ ( \nabla a_{nN} ) \times (\nabla a_{Nm} ) ] \nonumber \\ 
	&+ (- a_{mn} + a_{nm}) [ (\nabla a_{mN} ) \times (\nabla a_{Nn} ) - ( \nabla a_{nN} ) \times (\nabla a_{Nm} ) ] + 
	(\text{cyclic terms}) \bigl \} \nonumber \\ 
	=& 8i \sum_{m=1}^{N-2} \sum_{n=m+1}^{N-1} \bigl [ a_{mn} (\nabla a_{nN} ) \times (\nabla a_{Nm} ) + 
	a_{nm} (\nabla a_{mN} ) \times (\nabla a_{Nn} ) + (\text{cyclic terms}) \bigr ],  
\end{align} 
\end{widetext} 
which is equivalent to Eq.~(\ref{eq:s3}).

\end{document}